\documentclass[a4paper,11pt]{article}
\pdfoutput=1 
\usepackage{jcappub} 
\usepackage{amsmath}
\usepackage{amssymb}
\usepackage{float}
\usepackage[T1]{fontenc} 

\def\ds{{\rm d}s}

\newcommand{\Hydra}{\ensuremath{\tt Hydra}}
\newcommand{\healpix}{\ensuremath{\tt HEALPix}}
\newcommand{\camb}{\ensuremath{\tt CAMB}}
\newcommand{\smica}{\ensuremath{\tt SMICA}}
\newcommand{\nilc}{\ensuremath{\tt NILC}}
\newcommand{\sevem}{\ensuremath{\tt SEVEM}}
\newcommand{\commander}{\ensuremath{\tt Commander}}
\newcommand{\fwhm}{\ensuremath{\tt FWHM}}
\newcommand{\lcdm}{\ensuremath{\tt \Lambda CDM}}
\newcommand{\wmap}{\ensuremath{\tt WMAP}}
\newcommand{\planck}{\ensuremath{\tt Planck}}
\def\deg{\ifmmode^\circ\else$^\circ$\fi}
\def\pdeg{\ifmmode $\setbox0=\hbox{$^{\circ}$}\rlap{\hskip.11\wd0 .}$^{\circ}60\arcm
          \else \setbox0=\hbox{$^{\circ}$}\rlap{\hskip.11\wd0 .}$^{\circ}$\fi}
\def\arcs{\ifmmode {^{\scriptstyle\prime\prime}}
          \else $^{\scriptstyle\prime\prime}$\fi}
\def\arcm{\ifmmode {^{\scriptstyle\prime}}
          \else $^{\scriptstyle\prime}$\fi}
\newcommand{\Nside}{\ensuremath{N_{\rm side}}}
\newcommand{\ffp}{\ensuremath{\tt FFP8}}

\title{\boldmath Search for anomalous alignments of structures in \planck\ data using Minkowski Tensors}

\author[a,b,*]{Joby P. K.\note{* Corresponding author.}}
\author[a]{, Pravabati Chingangbam}
\author[c]{, Tuhin Ghosh}
\author[d]{, Vidhya Ganesan}
\author[b]{, Ravikumar C. D.}

\affiliation[a]{Indian Institute of Astrophysics, Koramangala II Block, \\Bangalore 560 034, India}
\affiliation[b]{Department of Physics, University of Calicut, \\Malappuram, Kerala 673 635, India}
\affiliation[c]{School of Physical Sciences, National Institute of Science Education and Research, HBNI, Jatni 752050, Odissa, India}
\affiliation[d]{Department of Physics, Indian Institute of Science, C. V. Raman Ave., \\Bangalore 560 012, India}

\emailAdd{joby@iiap.res.in}
\emailAdd{prava@iiap.res.in}

\abstract{Minkowski Tensors are tensorial generalizations of the scalar Minkowski Functionals. 
  Due to their tensorial nature they contain additional morphological information of structures, in particular about shape and alignment, in comparison to the scalar Minkowski functionals.  They have recently been used~\cite{Vidhya:2016} to study the statistical isotropy of temperature and $E$ mode data from the \planck\ satellite. The calculation in~\cite{Vidhya:2016} relied on stereographic projection of the fields to extract the shape and alignment information. 
  In this work, we calculate Minkowski Tensors directly on the sphere and compute the net alignment in the data, based on a recent work that extends the definition of Minkowski Tensors to random fields on curved spaces. This method circumvents numerical errors that can be introduced by the stereographic projection. We compare the resulting net alignment parameter values obtained from the frequency coadded CMB temperature data cleaned by the \smica\ pipeline\footnote{Based on observations obtained with \planck\ (\href{http://www.esa.int/Planck}{http://www.esa.int/Planck}), an ESA science mission with instruments and contributions directly funded by ESA Member States, NASA, and Canada.}, to those obtained from simulations that include instrumental beam effects and residual foreground and noise. We find very good agreement between the two within $\approx 1\sigma$. We further compare the alignments obtained from the beam-convolved CMB maps at individual \planck\ frequencies to those in the corresponding simulations. We find no significant difference between observed data and simulations across all \planck\ frequencies, except for the 30 GHz channel. For the 30 GHz channel we find $\approx 2 \sigma$ difference between the data and the simulations. This mild disagreement most likely originates from inaccurate estimation of the instrumental beam at 30 GHz in the  $\ensuremath{\tt FFP9}$ simulations.}

\begin{document}
\maketitle
\flushbottom
\section{Introduction}
\label{sec:intro}


An important tenet of the standard \lcdm\ cosmology is that the universe is statistically isotropic. However there is no fundamental reason why the universe should be statistically isotropic (and homogeneous) and hence testing this basic assumption is a crucial exercise in the construction of our understanding of the universe. Tests for isotropy rely on cosmological data such as the Cosmic Microwave Background (CMB) and the large-scale structure. The availability of high resolution data covering large fractions of the sky, particularly the CMB \cite{Penzias:1965,Smoot:1992,Bond:1987ub}, has resulted in high precision tests of statistical isotropy~\cite{WMAP9:2013,Adam:2015}. There have been reports of anomalous departures from statistical isotropy (SI) at large angular scales~\cite{Adam:2015}. Departures from SI in the CMB can be caused by physical processes in the generation and subsequent evolution of the primordial density fluctuations. Besides, even if the true CMB is SI, the observed CMB map could be anisotropic due anisotropy of the beam, data noise and other systematic effects related with the instrumentation. It is important to understand these effects for correct interpretation of the SI violation from the observed CMB map.


Several methods for testing the SI of CMB fields can be found in the literature, and one such method defined on harmonic space is the Bipolar Spherical Harmonics (BiPoSH). In \cite{Souradeep:2006}, the authors have demonstrated the use of the Bipolar power spectrum, as a test of SI. The condition for SI is that the Bipolar power spectrum must be zero for all multipoles other than the monopole. They have used \wmap\ 3-year data and find no evidence for violation of SI in the temperature fields, but find significant SI violation in the polarization maps. 
Another method for probing the SI of CMB fields is the power tensor method as demonstrated in \cite{Pranati:2017}. In this method, the CMB field is first decomposed into the usual spherical harmonics. The second rank power tensor is constructed from the products of the coefficients of the spherical harmonics. This power tensor has three eigenvalues and correspondingly three eigenvectors. The power entropy and alignment entropy are then defined in terms of the eigenvalues and the corresponding eigenvectors of the power tensor. The power entropy and the alignment entropy give a measure of SI of the field. Using this method, the authors find a very mild signal of statistical anisotropy in the \wmap\ 9-year ILC map, and the \planck\ component-separated CMB maps (\nilc, \sevem, and \smica).

Minkowski Tensors (MTs) \cite{McMullen:1997,Alesker:1999,Beisbart:2002,Hug:2008,Schroder3D:2013,Beisbart:2001a,Beisbart:2001b} are tensorial quantities that  generalize the scalar Minkowski Functionals (MFs) \cite{Tomita:1986,Gott:1990,Mecke:1994,Schmalzing:1997,Winitzki:1998,Matsubara:2003yt,Buchert:2017uup,COBE_NG:2000,WMAP_NG:2011,Ganesan:2014lqa,Ade:2015ava,Planckiso:2015,Chingangbam:2013,Coles:1987}. Recent applications of MTs to cosmological fields can be found in \cite{Vidhya:2016,mtsphere:2017,Stephen2d:2018,Stephen3d:2018,Akanksha:2017}. They have also been used to study the morphology of Galaxies in \cite{Shandarin:2003} and \cite{Shandarin:2004}. Out of the full set of MTs, we focus on the one that carries shape and alignment information of structures, and we refer to it as the Contour Minkowski Tensor (CMT). This particular MT can be used to test for SI of the CMB fields. Since MTs are defined on real space  they provide a complementary method of analysis to the ones mentioned above which are defined on harmonic space. MTs carry information of correlations of arbitrary order and hence can provide stringent constraints on SI violation. In \cite{Vidhya:2016}, the authors numerically computed MTs of the CMB temperature and polarization fields given in the \planck\ 2015 data release, to probe for SI violations in the fields. They found no significant violation of SI in the temperature fields, but they found $\approx 4\sigma$ deviation from SI in the \planck\ polarization data. Since MTs are defined on flat space, the calculations in~\cite{Vidhya:2016} relied on stereographic projection of the fields from the sphere to a plane. Stereographic projection does not alter shapes but introduces size scaling of structures, and hence has a non-trivial effect on the relative alignment between structure. 

The definition of MTs was generalized to random fields on smooth curved spaces, and a general numerical method for calculating the alignment parameter was developed, in~\cite{mtsphere:2017}. Using this method, in the present work, we calculate MTs directly on the sphere. This is done by converting the line integrals in the definition of MTs to area integrals over the entire map, thus getting rid of the need to compute boundary pixels. Also, since there is no stereographic projection involved, we need not worry about alterations in the shapes of the connected regions and holes. Hence, we expect our analysis to be accurate. In this work, we analyse only \planck\  temperature maps since the full polarization data is not yet released. We find no significant deviation from SI in the \smica\ CMB temperature map, as well as the \planck\ individual frequency beam-convolved CMB temperature maps in the frequency range, 44 to 857 GHz. However, we find $\approx 2\sigma$ difference between the data and the simulations for the 30 GHz beam-convolved CMB temperature map. We plan to extend this analysis to the CMB polarization fields when the full \planck\ data becomes available.


The paper is organized as follows. In section 2 we briefly review MTs on the sphere, focussing on the contour MT. In section 3 we test the method on CMB simulations and quantify SI. In section 4 we apply our method to \planck\ data and describe our results. We end with a summary of our results, discussion of their relevance and future directions in section 5.

\section{Contour Minkowski Tensor on the sphere}
\label{sec:mtonsphere}
MTs are in general defined to be $(m,n)$ ranked tensors on flat space, where $m,n$ are arbitrary positive integers. They get subdivided into translation invariant and translation covariant tensors. Here we follow the notation and generalization of MTs to curved space given in \cite{mtsphere:2017}. There are three rank-two MTs on the sphere, denoted by $\mathcal{W}_i$, $i=0,1,2$. Each of these corresponds to a tensor generalization of the three scalar MFs. The CMT is the tensorial counterpart of the scalar MF that gives the total contour length. For this work, we focus attention on the CMT as it can be used to test the SI of the observed CMB fields. In this section, we briefly describe the CMT and how it encodes information about alignment of structures.
\subsection{Definition of Contour Minkowski Tensor}
\label{sec:mtin2d}

\begin{figure}[!hb]
 \centering
 \resizebox{3.2in}{4.2in}{\includegraphics{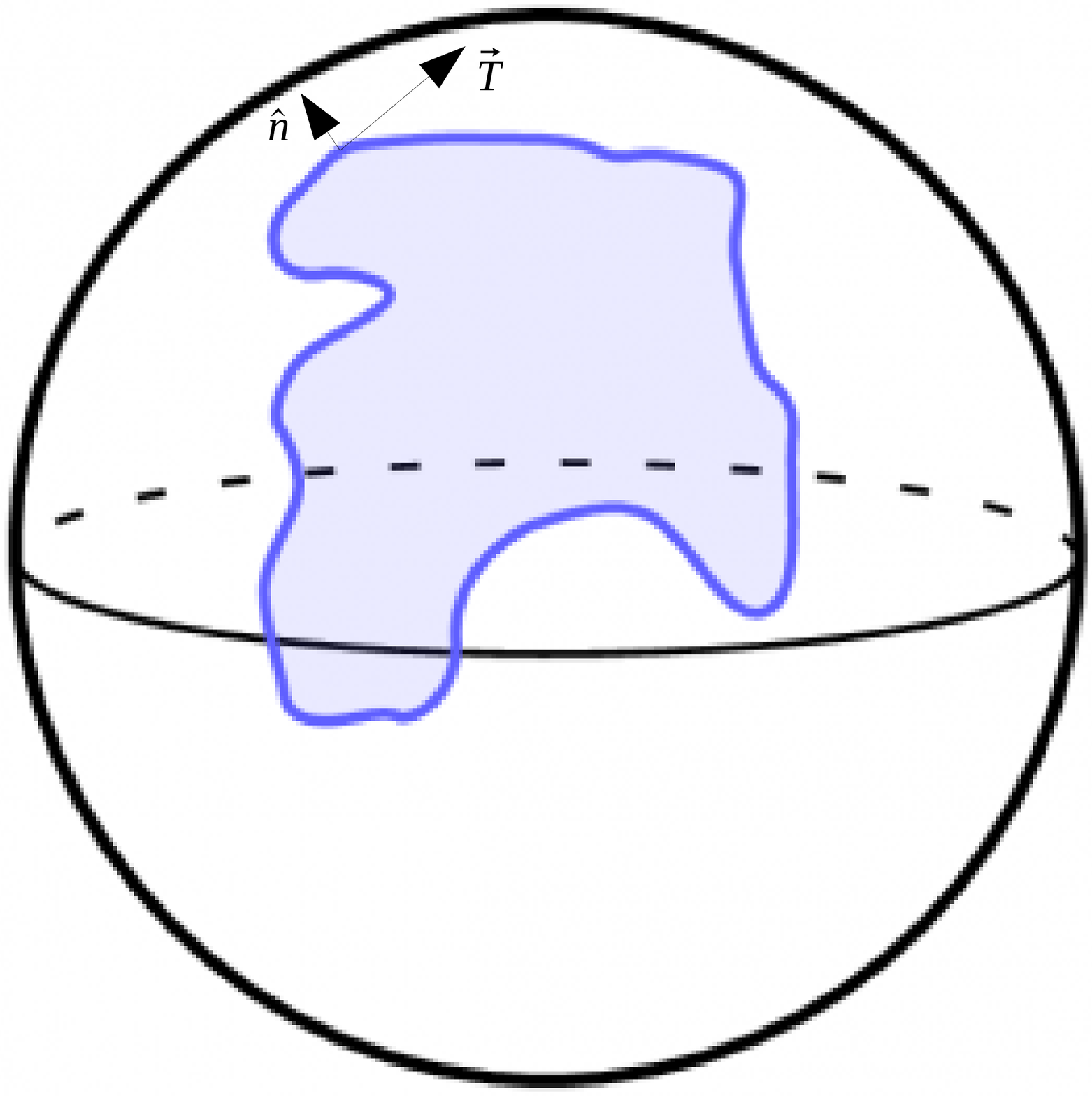}}
 \caption{\large{A closed curve on the surface of the unit sphere. $\hat{n}$ and $\vec{T}$ denote the unit normal to the curve and the tangent vector to the curve, respectively, at every point on the curve.}}
 \label{fig:ccurve}
\end{figure}

 Given a closed curve, $C$, on the sphere, the CMT denoted by $\mathcal{W}_1$, is defined as
\begin{equation}
 \mathcal{W}_1 = \frac14 \int_{C} \hat{T} \otimes \hat{T} \,\ds,\\
 \label{eqn:W1}
\end{equation}
where $(\hat T \otimes \hat T)_{ij}\equiv\frac12 \left( \hat T_i\hat T_j+ \hat T_j \hat T_i\right)$. 
The trace of $\mathcal{W}_1$ gives the total boundary length.

Given a random field on the sphere, the excursion set or level set associated with each threshold field value, $\nu_t$, gives a set of closed curves. Then the sum of $\mathcal{W}_1$ over all the curves will be given by
\begin{equation} 
 \overline{\mathcal{W}}_1(\nu_t) = \frac14 \int_{\mathcal S^2} \hat{T} \otimes \hat{T} \,\ds,\\
 \label{eqn:curlyw1}
\end{equation}
where the threshold dependence on the right hand side will be due to the unit tangent vectors.
\subsection{Probing SI violation using the CMT}
\label{sec:isotropy}
For a given closed curve in two dimensions, let $\lambda_1$ and $\lambda_2$, such that, $\lambda_1 \le \lambda_2$, denote the two eigenvalues of $\mathcal{W}_1$. Then, the shape parameter, $\beta$, is defined to be (see \cite{mtsphere:2017}), 
\begin{equation}
 \beta\equiv\frac{\lambda_1}{\lambda_2}.
 \label{eqn:beta}
\end{equation}
$\beta$ gives a measure of the isotropy of the closed curve. $\beta=1$ means that the curve is isotropic. For example, for a circle of any given finite radius, $\beta=1$. For curves that are not isotropic, the value of beta lies between 0 and 1. For an ellipse of semi-major axis, $a$, and semi-minor axis, $b$, larger the ratio $a/b$, smaller the value of $\beta$.

For a set of closed curves, we can sum the matrix $\mathcal{W}_1$ element by element to obtain $\overline{\mathcal{W}}_1$. Let $\Lambda_1$ and $\Lambda_2$ denote the eigenvalues of $\overline{\mathcal{W}}_1$ such that, $\Lambda_1 \le \Lambda_2$. Then the alignment parameter, $\alpha$, for the set of curves is given by,
\begin{equation}
 \alpha \equiv \frac{\Lambda_1}{\Lambda_2}.
 \label{eqn:alpha}
\end{equation}
$\alpha$ quantifies the statistical (an)isotropy of the set of closed curves. For example, for a pair of identical ellipses with semi-major axes aligned in the same direction, $0\le\alpha=\beta\le1$. However, for a pair of identical ellipses with their semi-major axes aligned in perpendicular directions, $\alpha=1$. While $\alpha$ does not capture the intrinsic isotropy of the individual curves, it carries information regarding their relative alignment. We can find the centroid of the curves and obtain in every direction, the mean radial distance of the points on the curves from the centroid. We can then construct the `locus curve' by connecting points in each direction that are at the mean distance in that direction from the centroid. We can then interpret $\alpha$ as the measure of isotropy of the `locus curve' (see \cite{mtsphere:2017}).

For a random field defined on the surface of a sphere, one can choose a threshold, $\nu_t$, for the field to obtain a set of connected regions and holes. The boundaries of these connected regions and holes form closed curves on the sphere. The isotropy and alignment of these curves contain information regarding the isotropy of the field. It is not practically possible to inspect them visually and infer the isotropy of the field since the shapes of the curves are arbitrary and the number of curves is usually very large. The alignment parameter $\alpha$ for a given map can be calculated numerically with a code, and hence provides a useful probe of the statistical isotropy of random fields.

 \subsection{Numerical computation of CMT on the sphere}
 \label{sec:computemt}
 Let $f$ denote a random field on the sphere for which we wish to compute the CMT. We first subtract the mean field value from the original field and divide it by the standard deviation to obtain the rescaled field, $u$, which by definition has zero mean and unit standard deviation. We then choose a threshold, $\nu_t$, of $u$, to obtain a set of connected regions and holes. The boundaries of these connected regions and holes form closed curves on the sphere and we can then compute the CMT as follows.
 
The line integral in Eq.~\eqref{eqn:curlyw1} can be transformed into an area integral by introducing a Jacobian, as done in \cite{Schmalzing:1998}, to give
 \begin{equation}
  \overline{\mathcal {W}}_1 = \frac14 \int_{{\mathcal S}^2} da \, \, \delta(u-\nu_t)\ |\nabla u| \ \hat{T} \otimes \hat{T},
  \label{eqn:w1random}
 \end{equation}
 where $\delta(u-\nu_t)$ is the Dirac delta function and $\hat{n}$ is the unit normal vector to the curve at every point on the curve. Now we need to express $\hat{T}$ in terms of the field. We know that for every point on the curve, the covariant derivative of the field on the sphere, $\nabla u$, is normal to the curve at that point. Here, $\nabla u = \left( u_{;1},u_{;2} \right)$, where $u_{;i}$ is the $i^{th}$ component of the covariant derivative of $u$. We can thus replace $\hat{n}$ with $\frac{\nabla u}{\left| \nabla u \right|}$. The components of the unit tangent vector can then be written as,
 \begin{equation}
  \hat{T}_i = \epsilon_{ij} \, \frac{u_{;j}}{\left| \nabla u \right|}.
  \label{eqn:hatt}
 \end{equation}
 Eq.~\eqref{eqn:w1random} can then be re-expressed as,
 \begin{equation}
  {\overline{\mathcal W}}_1 = \frac14\,\int_{{\mathcal S}^2}  da  \ \delta(u-\nu_t)\ \frac{1}{|\nabla u|} \ {\mathcal M},
  \label{eqn:w1final}
 \end{equation}
 where the matrix $\mathcal{M}$ is given by,
 \begin{equation}
  \mathcal M=  \left(
  \begin{array}{cc} 
    u_{;2}^2 &  u_{;1} \,u_{;2} \\
      u_{;1} \,u_{;2} & u_{;1}^2
  \end{array}\right).
  \label{eqn:M}
 \end{equation}
In order to numerically calculate Eq.~\eqref{eqn:w1final}, the $\delta-$function can be approximated as (\cite{Schmalzing:1998}),
 \begin{equation}
   \delta \left( u-\nu_t \right) = 
\left\{ \begin{array}{l}
    \frac{1}{\Delta \nu_t}, \quad {\rm if} \ 
   u \in \left( \nu_t-\frac{\Delta \nu_t}{2} , \nu_t+\frac{\Delta \nu_t}{2} \right)\\
    0, \quad {\rm otherwise},
  \end{array}
  \right.
  \label{eqn:delta}
 \end{equation}
 where $\Delta \nu_t$ is the bin size of the threshold values.

 The ensemble expectation value of $ {\overline{\mathcal W}}_1$ for Gaussian isotropic fields can be analytically calculated~\cite{mtsphere:2017}. It is given by
 \begin{equation}
 \langle {\overline{\mathcal W}}_1 \rangle_{ij} =   
\left\{ \begin{array}{l}
   \frac{1}{16 \sqrt{2} \, r_c} \, \exp(-\nu_t^2/2), \quad {\rm if} \ i=j  \\
   \\
    0, \quad i\ne j,
  \end{array}
  \right.
  \label{eqn:ensW1}
 \end{equation}
 where $r_c$ is the correlation length of the field and is given by $r_c=\sigma_0/\sigma_1$, where $\sigma_0=\sqrt{\langle f^2\rangle}$, and  {$\sigma_1=\sqrt{\langle|\nabla f|^2\rangle}$. From Eq.~\eqref{eqn:ensW1} we infer that $\alpha=1$ at all threshold values. For anisotropic fields the diagonal elements of ${\overline{\mathcal W}}_1$ will be unequal, Therefore, ${\overline{\mathcal W}}_1$ captures the anisotropy of a field by geometric quantification of the difference in the properties of the two components of the first derivative of the field.

   In practice, when we deal with finite resolution maps in a space of compact extent, such as the surface of a sphere, the number of curves at each threshold is finite (given by the Betti numbers~\cite{Chingangbam:2012,Park:2013}). As a consequence, we do not obtain $\alpha$ to be equal to exactly one at each threshold (see figure 4 of~\cite{mtsphere:2017}). $\alpha=1$ is recovered only in the limit of infinite resolution. At threshold values close to 0, the number of structures is the largest and hence $\alpha$ is closest to one. For a random distribution of a few curves, the probability that they will be isotropically distributed is very small. As a consequence, at higher $|\nu_t|$, the values of $\alpha$ decrease from one. It is, therefore, important to take into account the threshold dependence of $\alpha$ when searching for and interpreting results of SI on finite resolution fields.

   In \cite{Vidhya:2016}, the authors estimate $\alpha$ by using the projected CMB fields. First they stereographically project the CMB fields onto a plane and then an excursion set of this planar field is constructed for a given threshold. The Euclidean plane containing the field can be divided into area segments. Line segments are then assigned for each of these area segments based on whether the adjacent pixels belong to the excursion set or not. These line segments together form the boundaries of all the structures. The formula of MTs in pixelized space given in \cite{Schroder2D:2010} was used to finally estimate $\mathcal{W}_1$ and from it, $\alpha$. However the stereographic projection can distort the shapes of structures and lead to numerical errors in the computation of $\overline{\mathcal{W}}_1$ and $\alpha$.

The computation described here is done directly on the sphere. We therefore, expect more accurate estimation of $\overline{\mathcal{W}}_1$ and $\alpha$ by using this method. The data is in the form of pixelated maps and hence we substitute the $\delta$ function in \ref{eqn:w1final}, with the discrete $\delta$ function as in \cite{Schmalzing:1998}. This approximation of the $\delta-$function introduces a numerical error which can be analytically estimated~\cite{Lim:2012}. This error scales as the square of the threshold bin size, $\Delta \nu$, in the leading order.

\section{SI of simulated CMB maps}
\label{sec:simulations}

Before applying to observed \planck\ data, we first test the reliability of our method using simulated Gaussian CMB maps which have input isotropic power spectrum. We expect to recover $\alpha$ values close to one. As explained in section \ref{sec:computemt}, the numbers of hotspot and coldspot structures depends on the smoothing scale and the threshold level, and so $\alpha$ cannot be exactly one. We investigate the effect of Galactic and point sources mask, beam smoothing, and the pixel size on the estimation of $\alpha$ as a function of threshold.

\subsection{Effect of masking}
\label{sec:masking}
In regions close to the Galactic plane, the foreground contamination is very strong and hence it is difficult to reliably recover the clean CMB signal. These regions are therefore excluded in the cosmological analysis, along with the pixels that contain point sources. In our work, to remove these pixels from the analysis, we use the common mask UT78 for temperature analysis, which is given by the \planck\ team, as the preferred mask for temperature maps \cite{Adam:2015}. It has a sky fraction of 77.6\%. The process of masking removes a portion of the sky from the analysis. As a result, parts of the closed curves near the boundaries can get cut off and can contribute junk values to the calculation of $\alpha$. It is important to understand the impact of such effects on the calculation of the alignment parameter. We study the effect of masking the maps on the calculation of $\alpha$ with the help of simulations.

First, using \healpix\ \cite{Gorski:2005,HPX} and CMB angular power spectrum obtained from \camb\ \cite{Lewis:2000ah,cambsite}, we make 100 realizations of the CMB temperature sky with \fwhm=20\arcm\ and compute $\alpha$ for each of them. Then we compute the mean $\alpha$ over the 100 realizations and the standard deviation, $\sigma_{\alpha}$. The $\alpha$ thus obtained are represented in the top left panel of figure~\ref{fig:alpha_sim} by the blue line. We observe that the values of $\alpha$ for thresholds further away from the mean, are relatively smaller. For these thresholds, $\alpha$ is averaged over fewer structures and there is a very small probability of them being isotropically distributed. Thus they have smaller $\alpha$ values and agree well with the explanation in section \ref{sec:computemt}. Next, we mask the 100 CMB temperature realizations with UT78. We then repeat the procedure mentioned above, for the masked simulations and compute $\alpha$ for the masked maps. These are represented by the red line in the top left panel of figure~\ref{fig:alpha_sim}. We find that $\alpha$ values for the masked simulations are smaller than those for the unmasked ones. This is expected since masking removes a portion of the sky from the analysis, and hence leads to fewer structures that $\alpha$ is averaged over, and hence a larger statistical error.

\subsection{Effect of smoothing}
\label{sec:smoothing}
The observed CMB data contains residual noise from the observing instrument and this can lead to an inaccurate estimation of $\alpha$. We therefore smooth the maps by a Gaussian beam with an appropriate \fwhm\ to make sure that the effect of noise on the calculation of $\alpha$ is minimal. Also, as mentioned in section \ref{sec:masking}, the masking procedure can cut off closed curves near the boundaries of the mask and can contribute junk values to the $\alpha$ calculation. Smoothing the mask provides a way to mitigate this effect. 

To demonstrate the effect of smoothing the maps on the estimation of $\alpha$, we repeat the procedure mentioned in section \ref{sec:masking} with a different smoothing angle of 60\arcm, and calculate $\alpha$ for the masked simulations. These are represented by the red line in the top right panel of figure~\ref{fig:alpha_sim}. The blue line represents the masked simulations with a smoothing \fwhm\ of 20\arcm. We find that $\alpha$ values are relatively smaller for maps smoothed to \fwhm=60\arcm\ compared to those for maps smoothed to \fwhm=20\arcm. This is expected as smoothing reduces the number of structures and hence leads to a larger statistical error in the computation of $\alpha$.

\subsection{Effect of pixel size}
\label{sec:pixelsize}
The process of smoothing a map involves decomposing it into spherical harmonics and convolving the coefficients of the spherical harmonics with a Gaussian beam of the specified value of the \fwhm. It is also important to make sure that the smoothing is done over a small number of pixels. If we are smoothing the maps to a given \fwhm, then the maps should have a resolution parameter, \Nside, such that the smoothing scale is not too large as compared to the size of a pixel. We computed $\alpha$ for the SMICA CMB temperature data and for 100 pure CMB simulations, each having resolution Nside=2048, with and without downgrading the maps to Nside=512. We found that downgrading the maps leads to a better agreement between the data and the simulations, the reason for which is not well understood. To demonstrate the effect of pixel size, we first compute $\alpha^{\rm sim}$ for 100 pure CMB simulations with a resolution of \Nside=1024. We then compare that with $\alpha^{\rm sim}$ computed from the same simulations after first downgrading them to a resolution of \Nside=512. In both the cases, the maps are smoothed with a Gaussian beam of \fwhm=20\arcm. These are represented in bottom middle panel of figure~\ref{fig:alpha_sim} by the blue and red lines respectively. We find that, for the downgraded maps, the $\alpha$ values are smaller than the high \Nside\ maps. The process of downgrading involves storing the mean of the field values of four adjacent pixels, into one pixel in the downgraded map. This is similar to the effect of beam smoothing and leads to fewer structures in the map that $\alpha$ is averaged over. Thus we expect that downgrading the maps before calculation of $\alpha^{\rm sim}$ should lead to smaller $\alpha^{\rm sim}$ values.

\begin{figure}
  \centering
  \resizebox{3.1in}{2.2in}{\includegraphics{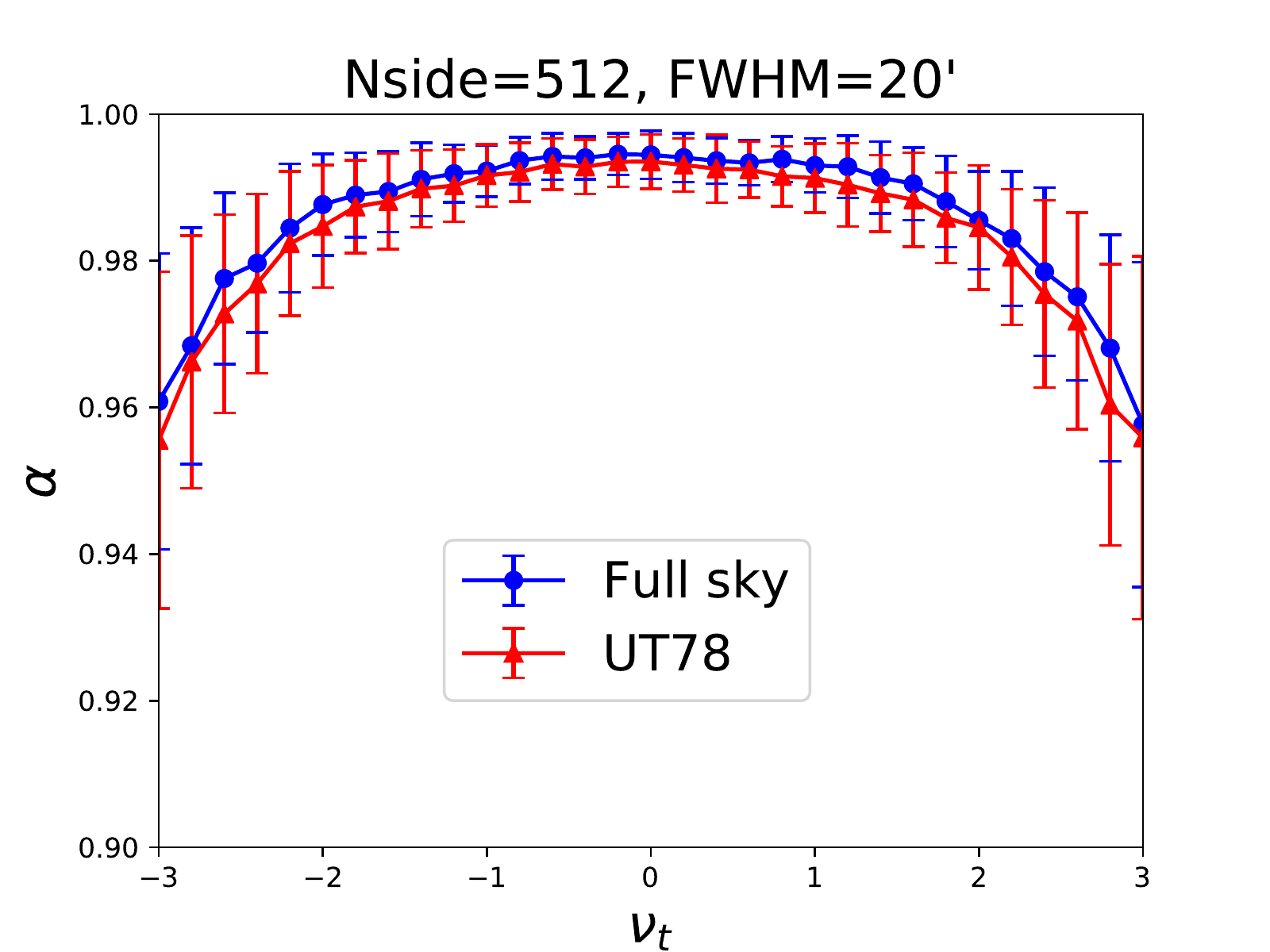}}
  \hskip -0.5cm \resizebox{3.1in}{2.2in}{\includegraphics{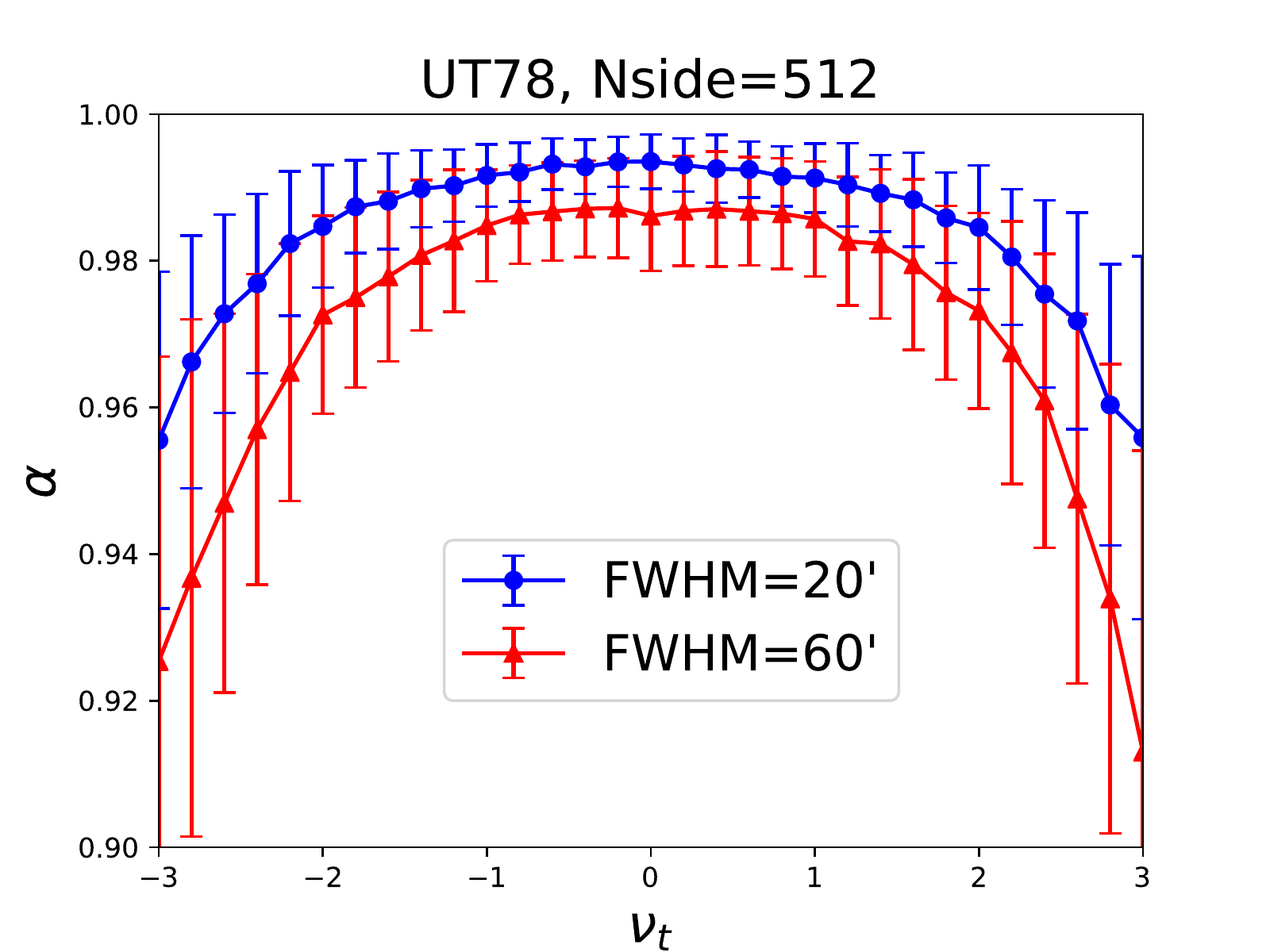}}
   \resizebox{3.1in}{2.2in}{\includegraphics{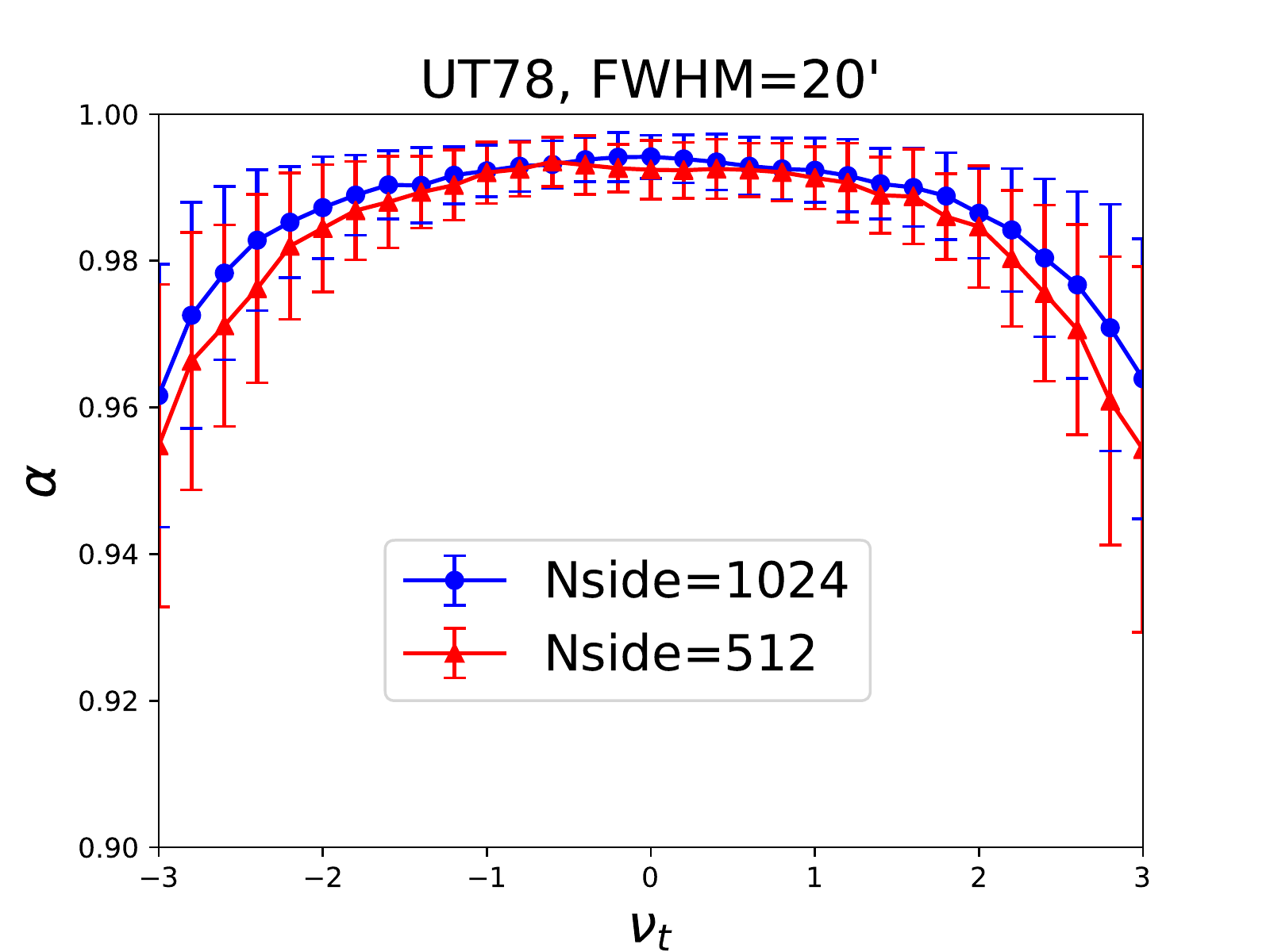}}
\caption{{\em Top left panel}: Comparison of $\alpha^{\rm sim}$ for simulated CMB temperature maps, smoothed with \fwhm=$20\arcm$, for unmasked (blue line) and masked (red line) simulations. {\em Top right panel}: Comparison of $\alpha^{\rm sim}$ for masked CMB temperature simulations, with smoothing \fwhm=20\arcm\ (blue line) and \fwhm=60\arcm\ (red line).  {\em Bottom panel}: Comparison of $\alpha^{\rm sim}$ values computed from CMB temperature simulations having \Nside=1024 and \fwhm=20\arcm, with and without first downgrading the maps to \Nside=512. All plots are obtained by averaging over 100 realizations. The error bars are the sample variance for 100 realizations. }
\label{fig:alpha_sim}
\end{figure}

\section{Toy examples of anisotropic fields and $\alpha$ estimation}
Before applying the CMT analysis to look for SI violation in real data, it is important to know the effect of anisotropy in the fields, on $\alpha$. The foreground and noise maps are not isotropic and provide a means to study how the anisotropy in the fields, is reflected in the $\alpha$ calculated from them. In this section, we discuss toy models of anisotropic fields, based on CMB foregrounds and noise and calculate $\alpha^{\rm sim}$ for such fields.

\subsection{Simulated maps with residual foreground contamination}
  We simulate 100 Gaussian sky realizations of CMB with resolution parameter \Nside=1024. Then to each realization, we add a fraction of the foreground map at 70 GHz provided by the \commander\ cleaning method, given in \cite{PLA} as the CMB subtracted maps. Let the CMB field be denoted by, $u$, and the foreground field be denoted by, $fg$. Then then final field, $f$, is obtained by adding the two fields: $f=u+\epsilon * fg$, where $\epsilon$ is the parameter that determines the level of foreground contamination. The foreground maps are not isotropic and hence the resultant maps contain a certain level of anisotropy. We calculate $\alpha$ for each of these maps and obtain $\alpha^{\rm sim}$, the mean $\alpha$ over all the 100 maps. 
  
  The $\alpha^{\rm sim}$ for the maps with varying levels of residual foreground contamination are shown in figure~\ref{fig:cmb_fg}. The black line represents maps with CMB-only simulations. The blue, green, and red lines represent maps with $\epsilon =$ 0.5, 1 and 2 respectively. We observe that for threshold values close to 0 and for negative threshold values, the impact of foreground contamination on $\alpha^{\rm sim}$ is very small. However, for thresholds greater than 1, there is a steady decline in $\alpha^{\rm sim}$ with increasing levels of foreground contamination.
  
  This effect can be explained as follows. After applying UT78, the galactic region of the map is removed from the analysis. In the remaining regions, the variance of the CMB fluctuations is much greater than that of the given foreground. Also, the mean of the CMB field is close to zero, while the mean of the foreground field is a relatively large positive number. The CMB field has negative as well as positive values, while the foreground field values are strictly positive. Thus, when the CMB and foreground fields are added pixel by pixel and the mean is subtracted from the resultant field, the CMB field dominates at thresholds smaller than the mean, while the foreground field dominates at larger thresholds. Hence, $\alpha^{\rm sim}$ values for negative thresholds are not strongly affected with increasing levels of foreground contamination, while the effect of foreground contamination is very apparent at larger thresholds.

\begin{figure}
  \centering
  \resizebox{3.1in}{2.2in}{\includegraphics{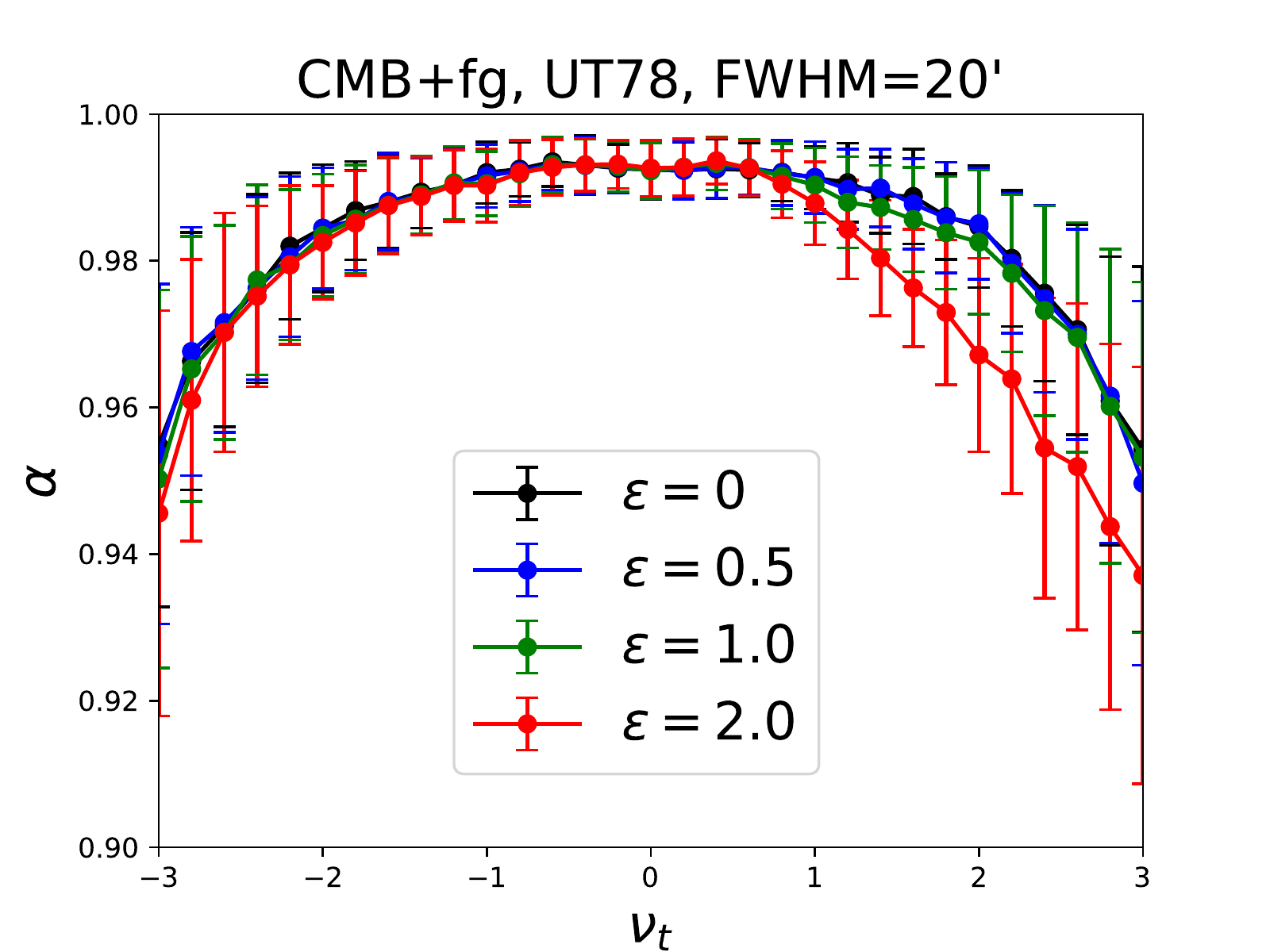}}
  \hskip -0.5cm \resizebox{3.1in}{2.2in}{\includegraphics{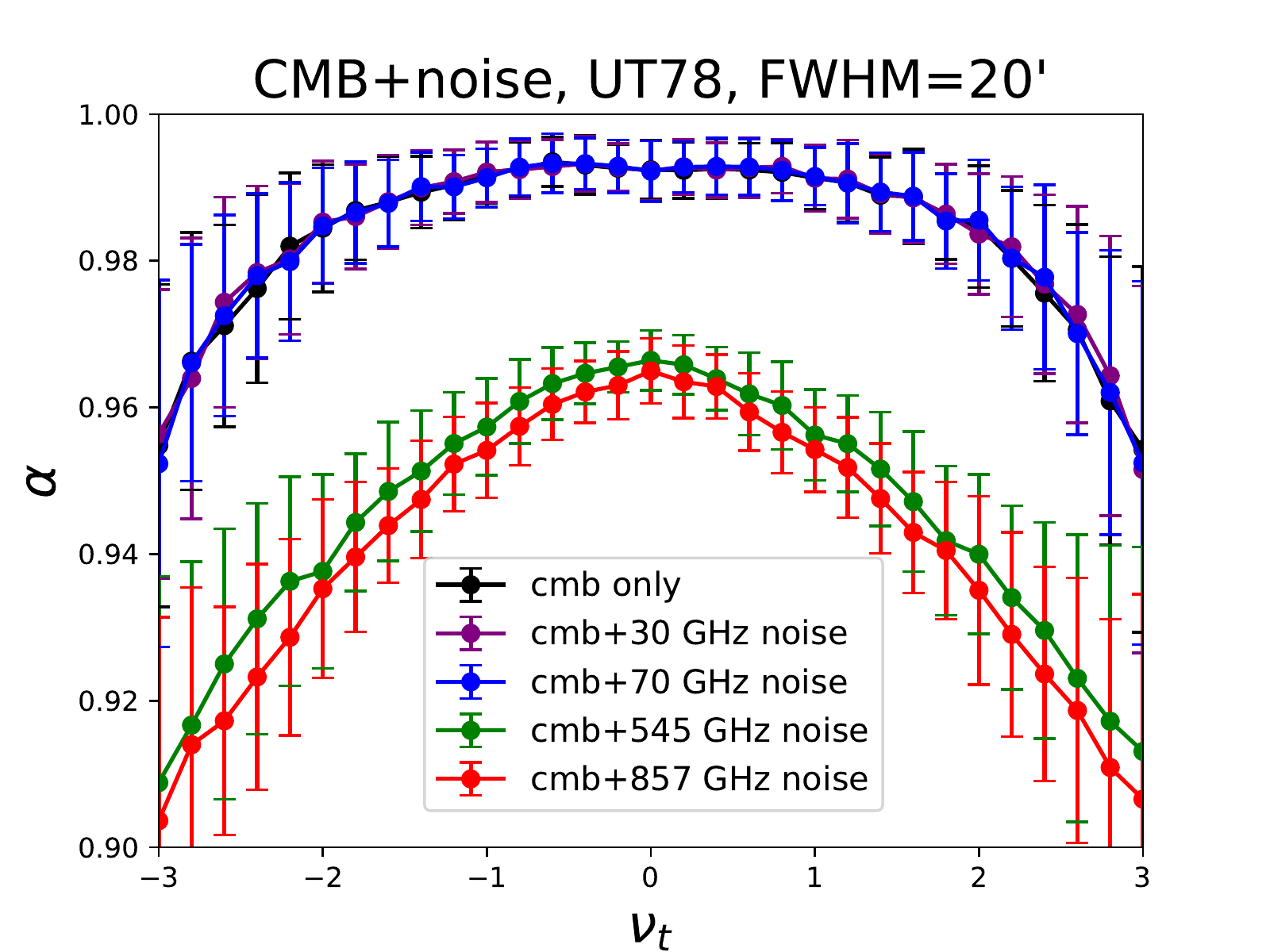}}
  \caption{{\em Left panel}: $\alpha^{\rm sim}$ for isotropic Gaussian simulated maps with some added foreground fraction. {\em Right panel}: $\alpha^{\rm sim}$ for isotropic Gaussian simulated maps with added \planck\ noise.} 
  \label{fig:cmb_fg}
\end{figure}

\subsection{Simulated maps with \planck\ noise added}
  Here we calculate $\alpha^{\rm sim}$ for simulated CMB maps to which \planck\ noise, given in \cite{PLA} has been added. First we simulate 100 Gaussian and isotropic realizations of the CMB temperature sky and then to each realization, we add the 70 GHz \planck\ noise simulation maps, pixel by pixel. We then compute $\alpha^{\rm sim}$, the mean $\alpha$ over all the 100 simulations. Let $u$ denote the CMB field and $n$ denote the noise field. Then the final field, $f$, is the addition of $u$ and $n$, $f = u + n$. The noise maps are not isotropic and hence there is net anisotropy in the final CMB+noise maps.
  
  $\alpha^{\rm sim}$ for simulated maps with added \planck\ noise is shown in the right panel of figure~\ref{fig:cmb_fg}. The black line represents maps that contain only CMB simulations. The purple, blue, green and red lines represent maps with CMB simulations and added \planck\ 30, 70, 545 and 857 GHz noise simulations respectively. We find very little deviation from SI in the case of the 30 and 70 GHz channels. However, there is significant deviation from SI in the CMB+noise maps corresponding to the noise at 545 and 857 GHz. This is expected because the noise estimates for the 545 and 857 GHz channels are very high, unlike the 30 and 70 GHz channels where the noise levels are much lower.

\section{Application of CMT to \planck\ data}
\label{sec:application}

  In this section, we apply the method of calculation of the CMT described in section \ref{sec:computemt}, to \planck\ data. Our goal is to probe the SI of the data by searching for disagreements  of the values of $\alpha$ computed from observed data and mock \planck\ data. Both the data and the simulations are masked with UT78, downgraded to Nside=512 and smoothed with a Gaussian beam of FWHM=20', before computing $\alpha$ from them. The numerical error arising from the approximation of the $\delta$ function mentioned in section \ref{sec:computemt}, is well understood \cite{Lim:2012} and is almost identical for different realizations having the same resolution. Hence, in our analysis, we have not subtracted this error since it doesn't affect the difference between the $\alpha$ values computed from the data and the simulations. In all the figures, both the $\alpha$ values for the data and $\alpha^{\rm sim}$ values for simulations are denoted by $\alpha$.
  
\subsection{$\alpha$ for \smica\ CMB temperature map}
\label{sec:planck} For our analysis, we use the CMB temperature map provided by the \smica\ pipeline \cite{PLA}. This map has \Nside=2048 and \fwhm=5\arcm. We first downgrade the map to a resolution of \Nside=512, smooth to an effective \fwhm\ of 20\arcm, and mask it with the common mask UT78 (effective sky coverage of 77.6\%). We then compute $\alpha^{\rm data}$ from this map, it is represented by the red line in figure~\ref{fig:smica}. Next, we take 100 publicly available Full Focal Plane (\ffp) \smica\ simulated maps which include instrumental beam effect and residual foreground and noise, and downgrade them to a resolution of \Nside=512 and  smooth them to an \fwhm\ of 20\arcm. We then mask them with UT78 and compute $\alpha^{\rm sim}$ from them. This is represented by the black line in figure~\ref{fig:smica}. We compare the data, which contain the CMB, instrumental effects and residual foreground and noise, with the \ffp\ \smica\ simulations which contain the same components. Based on our analysis, we find an excellent agreement between the data and the simulations, $\leq 1\sigma$. We reach the same conclusion when we compare $\alpha^{\rm data}$ with $\alpha^{\rm sim}$ computed from 100 CMB-only simulations. 

\subsection{$\alpha$ for individual \planck\ frequency channel CMB maps}\label{sec:LHFI} 
Next we look for alignment of structures in the beam-convolved CMB maps at individual frequencies. \planck\ provides observed sky temperature maps in nine frequency channels (see \cite{PLA}), ranging from 30 to 857 GHz, which we shall denote by, $T^{\rm obs}$. There are four main cleaning methods, namely, \commander, \nilc, \sevem, and \smica, which provide CMB-subtracted maps, for each of the frequency channels, which we denote by, $T^{\rm fg}$. We then get the cleaned beam-convolved CMB maps, $T^{\rm clean}$, in each frequency channel, by subtracting the foreground maps from the observed sky temperature maps, as
\begin{equation}
T^{\rm clean}\equiv T^{\rm obs}-T^{\rm fg}.
\end{equation}

 The LFI data has a resolution of \Nside=1024, while the HFI data are given at a resolution of \Nside=2048. In this work, all the maps are first downgraded to a resolution of \Nside=512, masked with UT78, and smoothed to an \fwhm\ of 20\arcm, before computing MTs from them. We compute $\alpha$ for these data and simulations. The results for individual frequency channels are plotted in figure~\ref{fig:alpha_LFI}.

In order to interpret the SI of the observed data described above, we need to compare them with mock \planck\ data.  For this purpose, we use 100 $\ensuremath{\tt FFP9}$ simulations for the individual frequency maps, of the CMB temperature sky given in \cite{PLA}. We compare the data, which contains CMB and instrumental beam effects, with the $\ensuremath{\tt FFP9}$ simulations containing the same. In figure~\ref{fig:alpha_LFI}, the black lines represent the $\ensuremath{\tt FFP9}$ simulations, while the purple, blue, green and red lines represent the maps cleaned by the \commander, \nilc, \sevem, and \smica\ pipelines, respectively. Again $\alpha^{\rm sim}$ is the mean over all the 100 simulations, and $\sigma_{\alpha}$ is the standard deviation. $\sigma_{\alpha}$ serves as the $1\sigma$ error bar for the purpose of our calculations.

\begin{figure}
 \centering
 \resizebox{3.1in}{2.2in}{\includegraphics{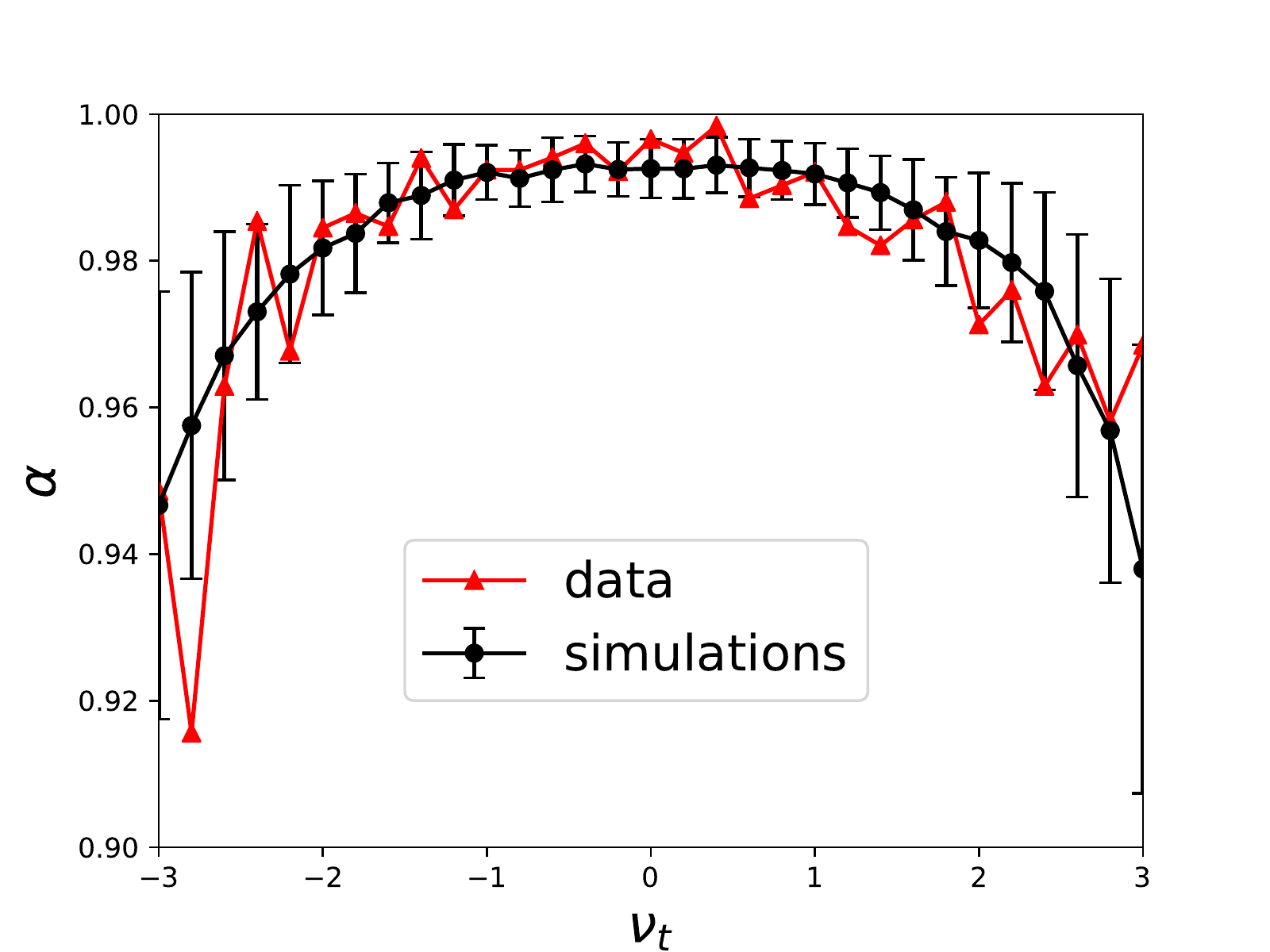}}
 \caption{Comparison of $\alpha$ computed from the \smica\ CMB temperature map (red line), and $\alpha^{\rm sim}$ computed from 100 \ffp\ simulations (black line).}
 \label{fig:smica}
\end{figure}

As explained in section \ref{sec:simulations}, the $\alpha$ values are smaller for thresholds $\nu_t$ further away from 0. For the individual frequency channels, we find that for the range 44 to 857 GHz, the $\alpha$ calculated from the data is consistent ($\approx 1\sigma$) with that calculated from the simulations. However, for the 30 GHz channel, the data deviates significantly from the simulations. In the case of the 30 GHz channel, the disagreement between data and simulations is at the level of $2\sigma$. The reason for this disagreement is not yet understood, but we suspect that it might be caused by an inaccurate estimation of the beam effect in the maps. Further investigation of the dependence of $\alpha$ on the instrument beam effect is required to confirm the source of this disagreement. Additionally, we repeat the same set of calculations for each frequency channel after smoothing the maps to an \fwhm\ of 60\arcm\ and find the same trend in the results. We also find that the addition of noise to the simulations, does not strongly affect the estimation of $\alpha$ for frequencies up to 353 GHz. However, for the 545 and 857 GHz channels, the addition of noise drastically changes the $\alpha$ estimates for the simulations. This result is due to the known fact that the CMB signal to noise ratio is smaller for the 545 and 857 GHz observation channels.

\begin{figure}
    \centering
    \resizebox{1.1in}{0.8in}{\includegraphics{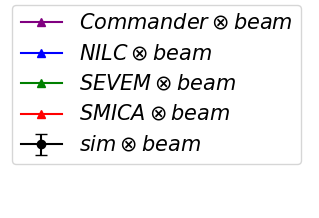}}
    
    \vskip 0.4cm
    \resizebox{1.9in}{1.2in}{\includegraphics{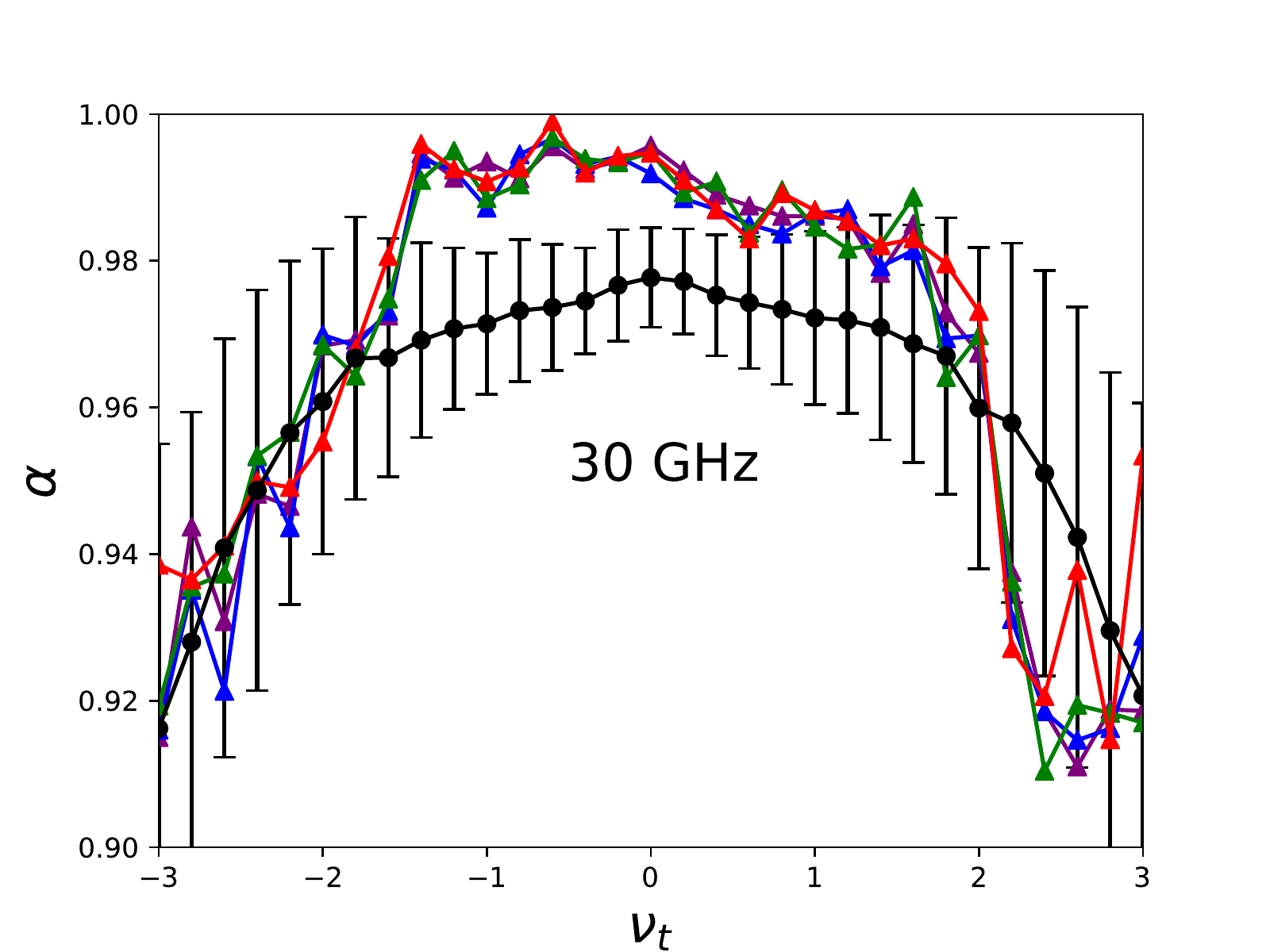}}
    \resizebox{1.9in}{1.2in}{\includegraphics{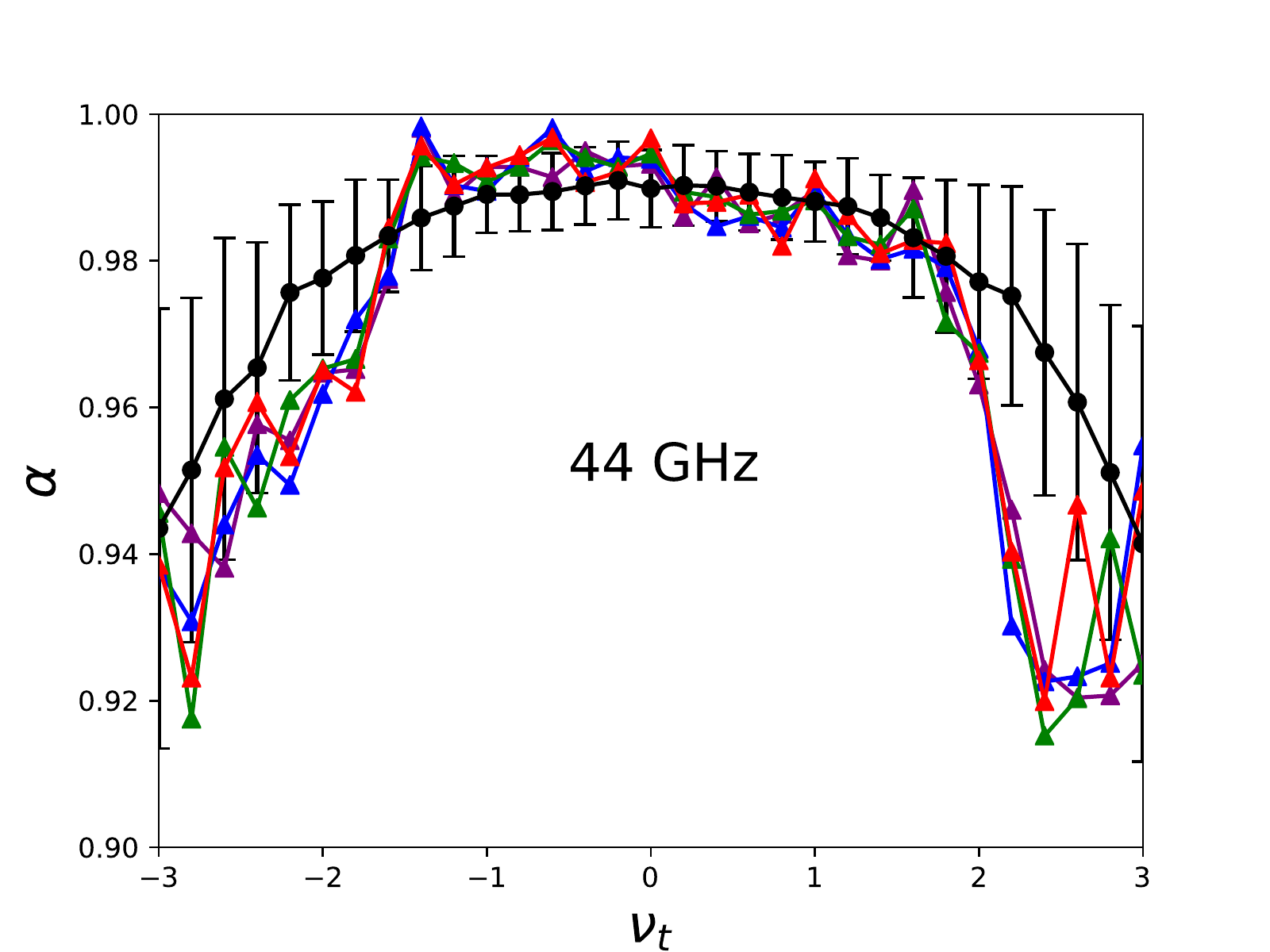}}
    \resizebox{1.9in}{1.2in}{\includegraphics{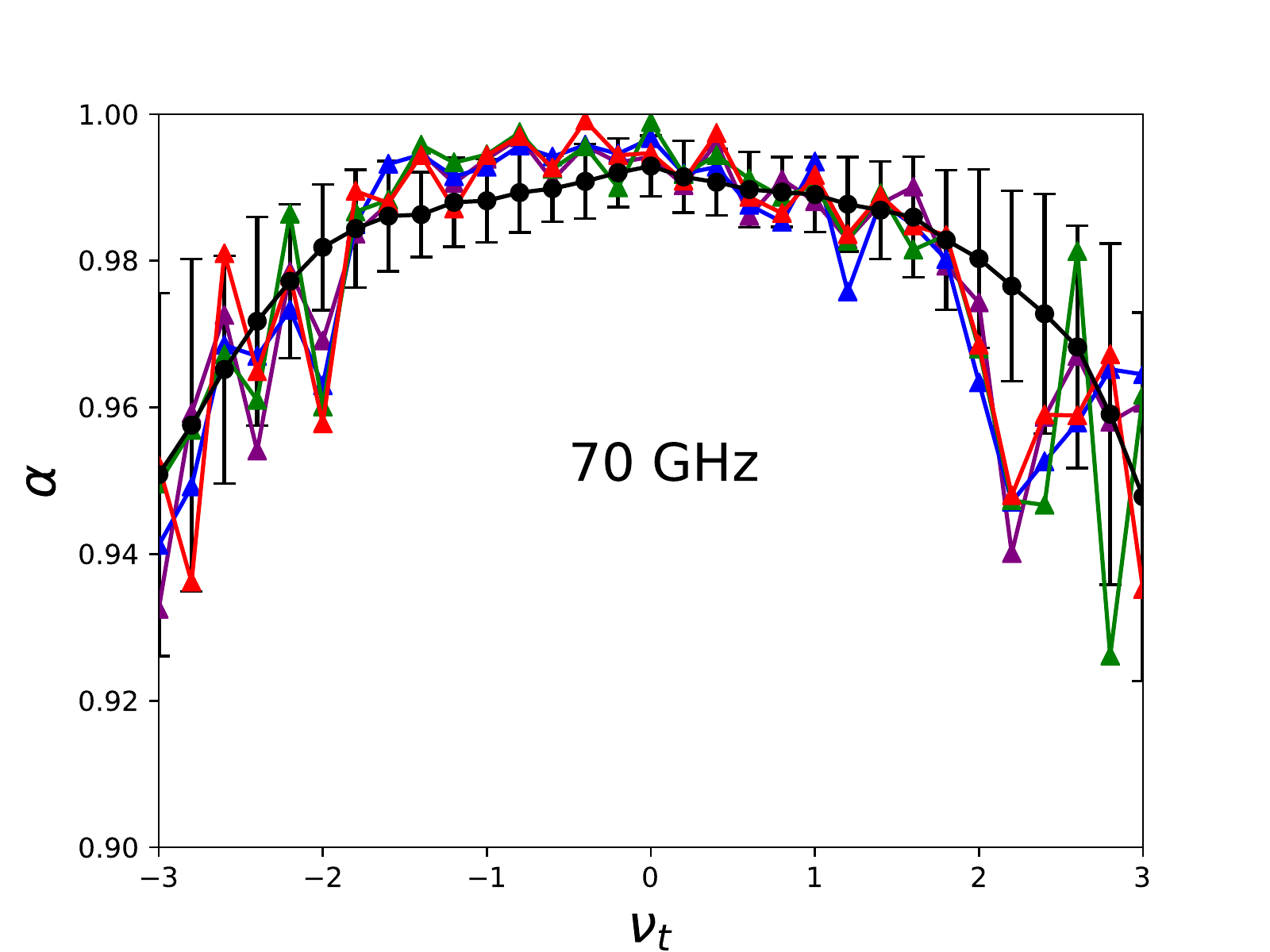}}
    
    \vskip 0.4cm
    \resizebox{1.9in}{1.2in}{\includegraphics{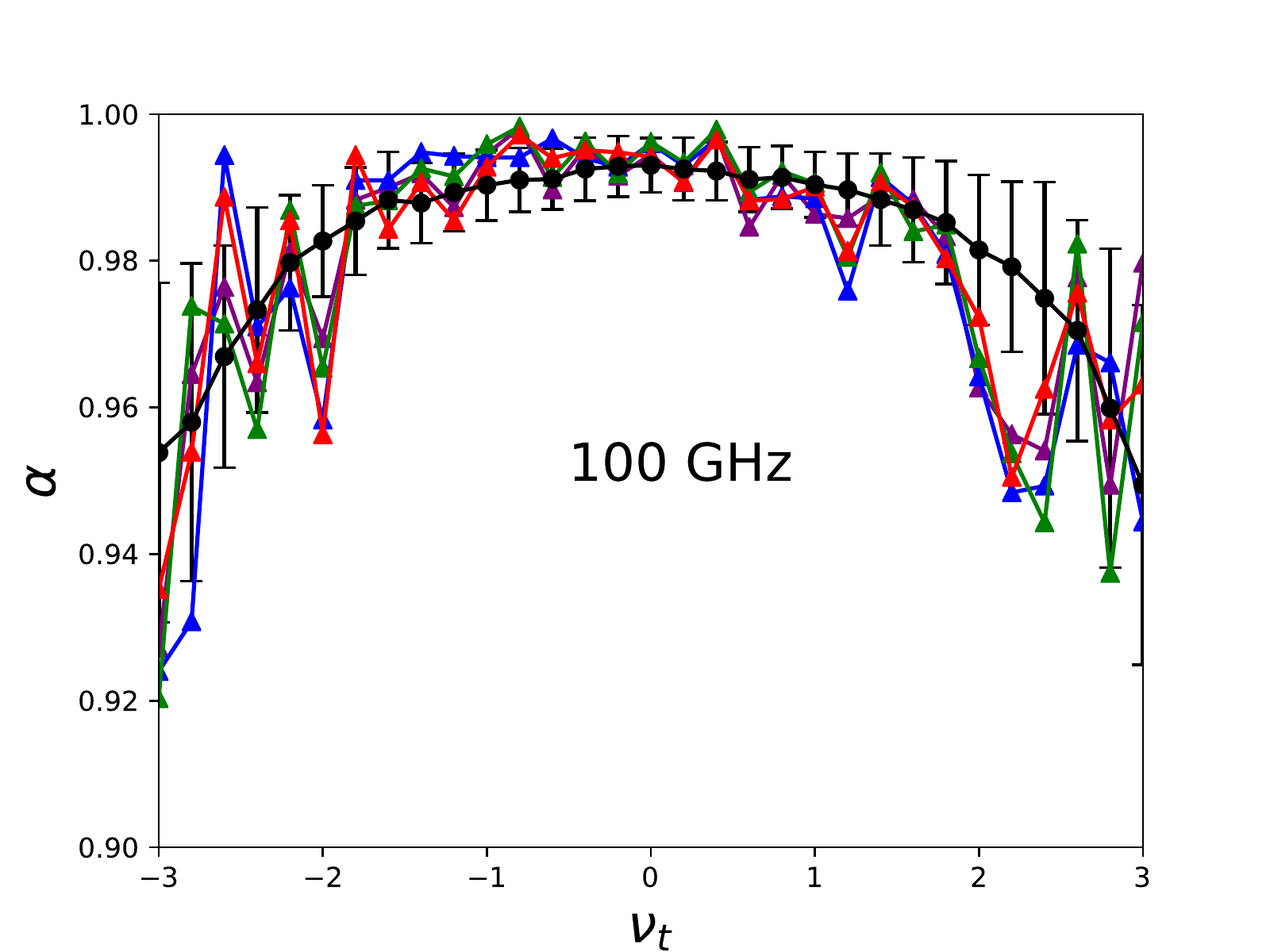}}
    \resizebox{1.9in}{1.2in}{\includegraphics{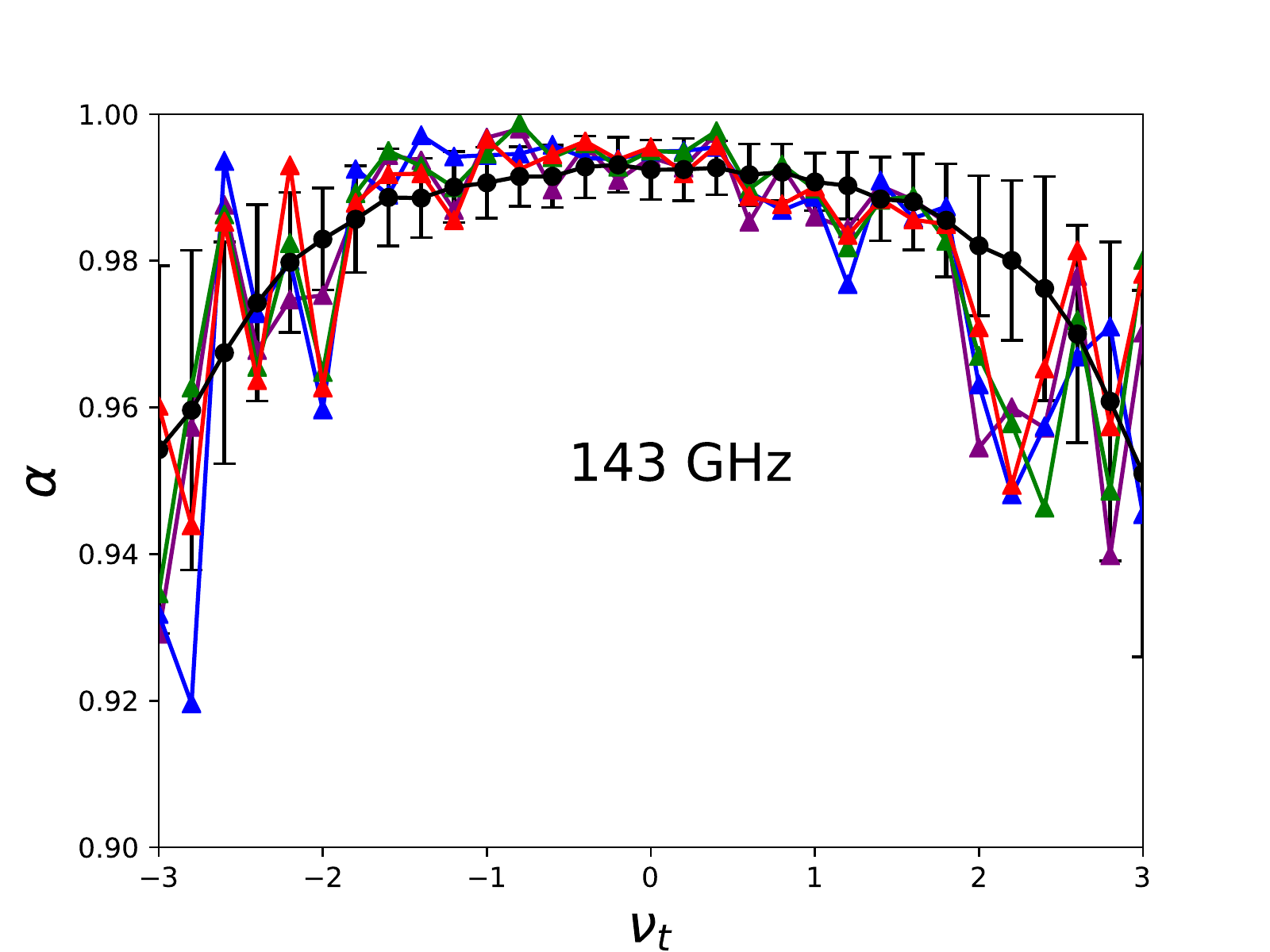}}
    \resizebox{1.9in}{1.2in}{\includegraphics{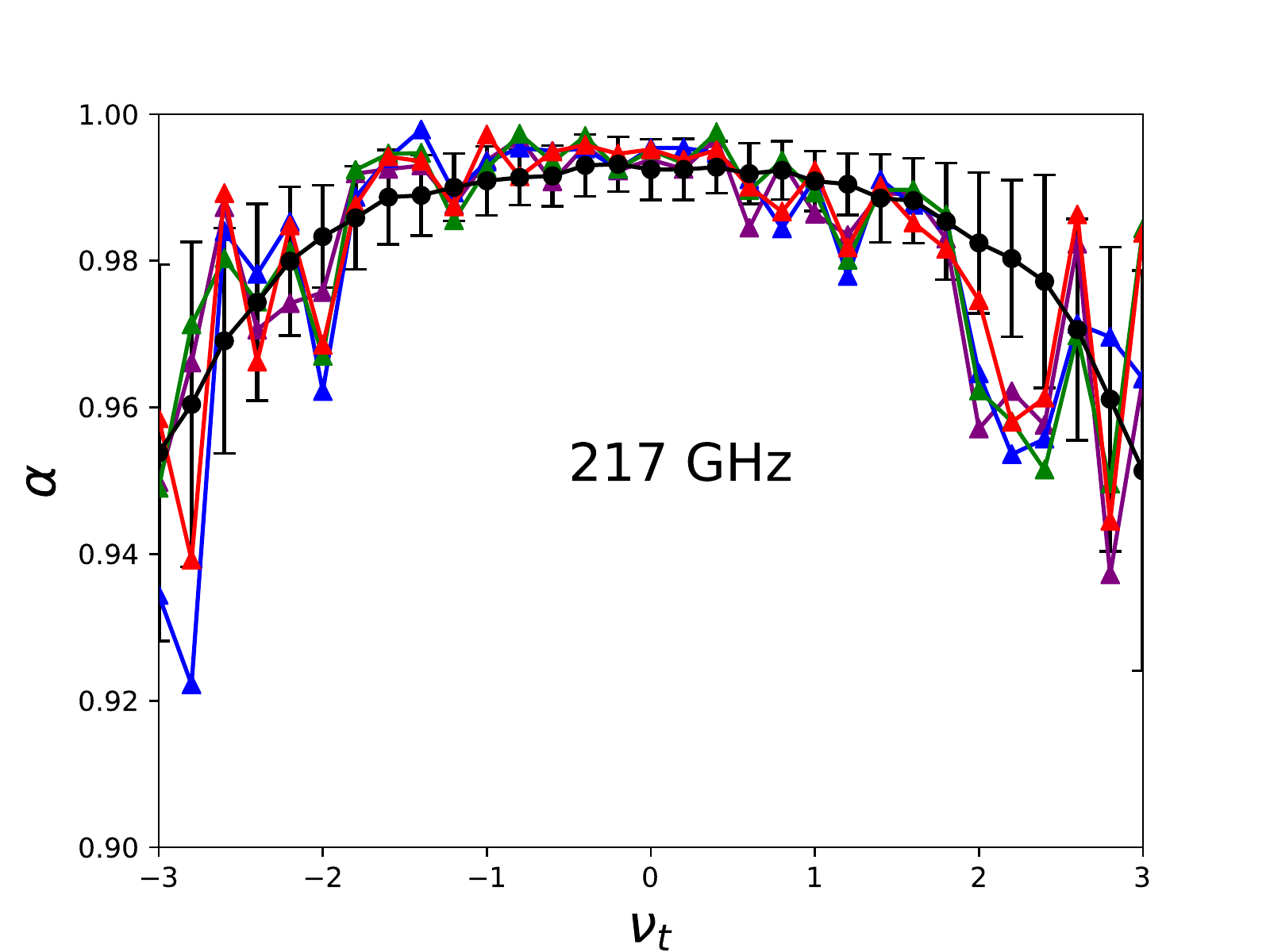}}
    
    \vskip 0.4cm
    \resizebox{1.9in}{1.2in}{\includegraphics{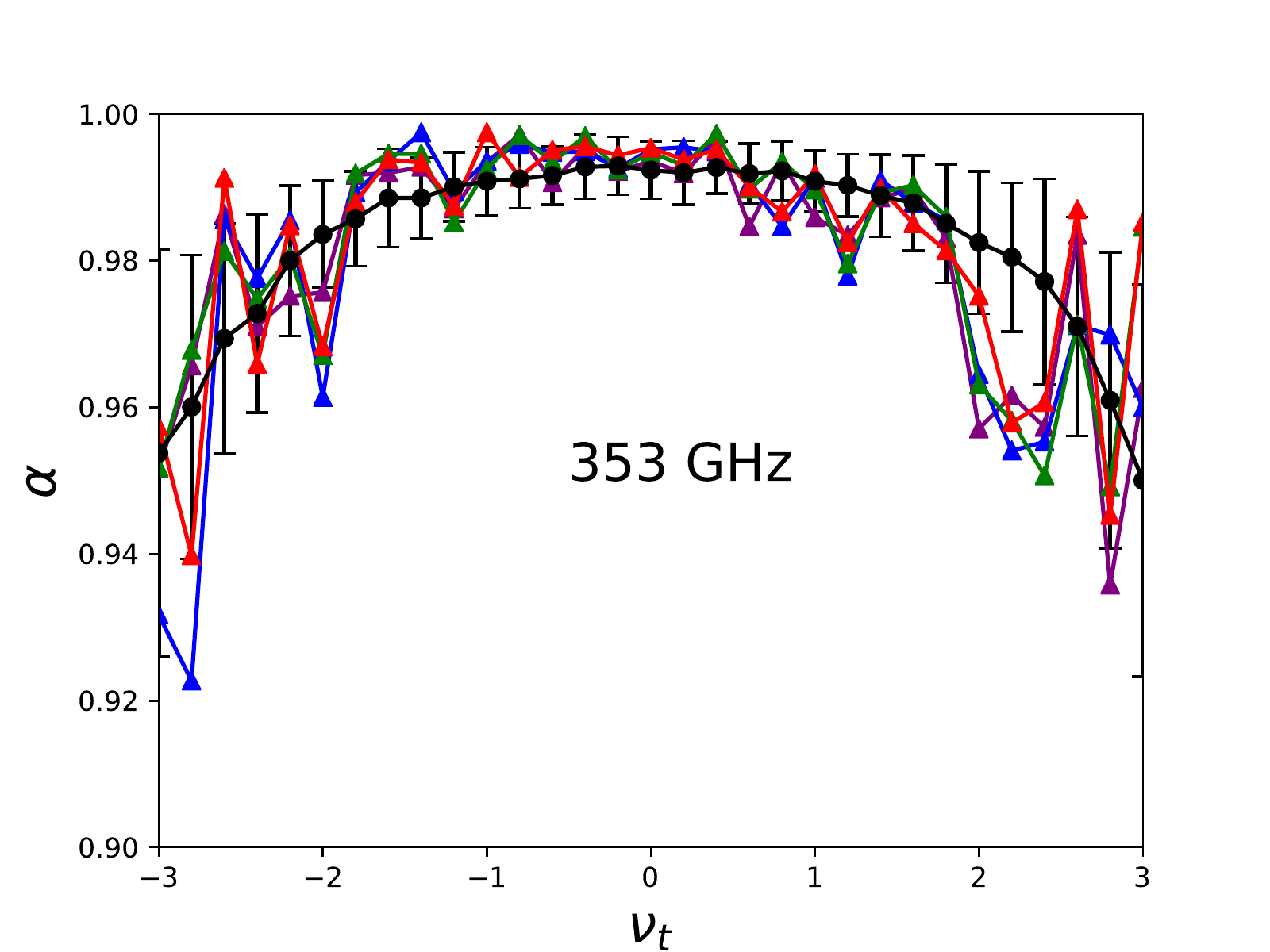}}
    \resizebox{1.9in}{1.2in}{\includegraphics{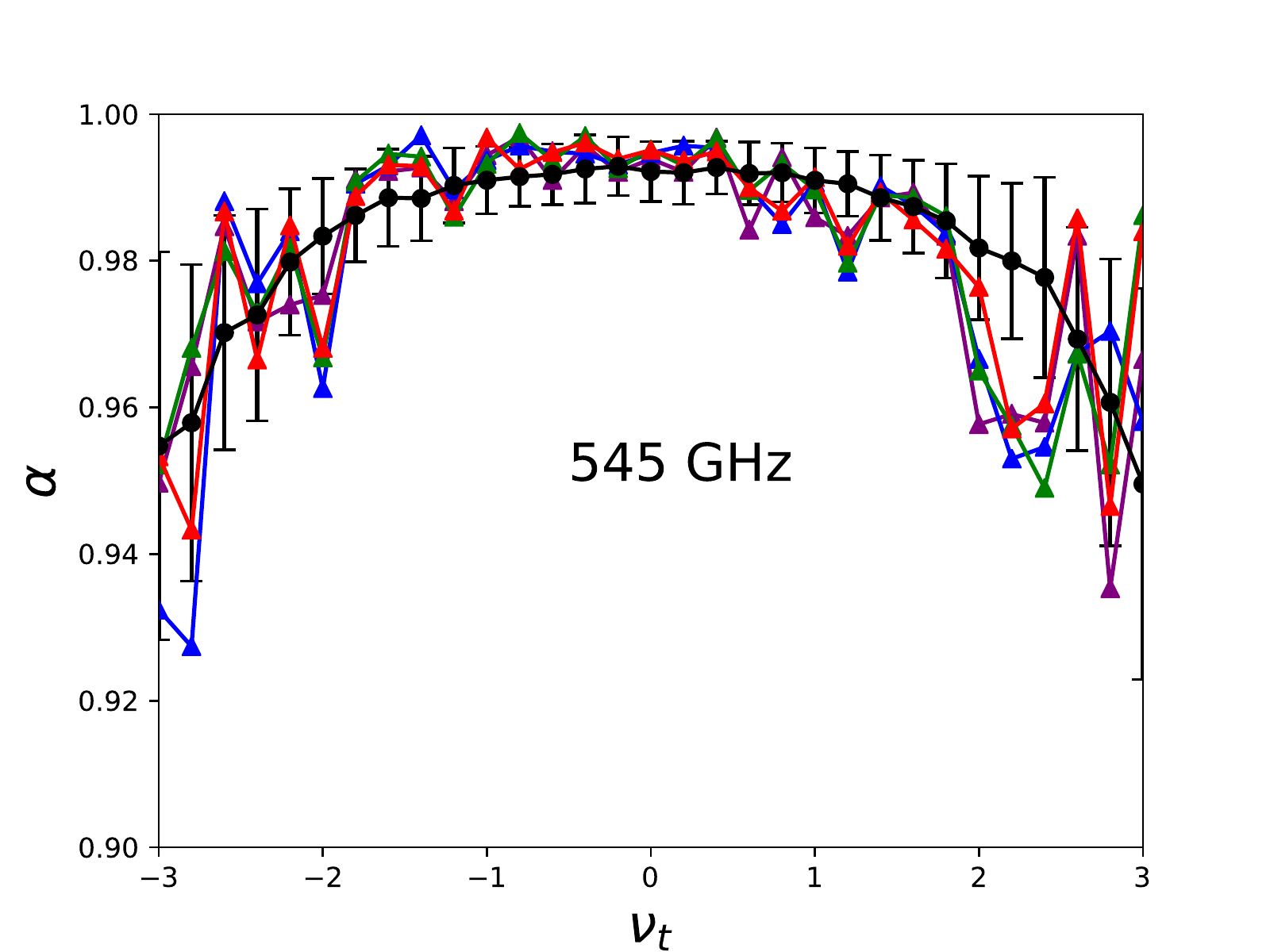}}
    \resizebox{1.9in}{1.2in}{\includegraphics{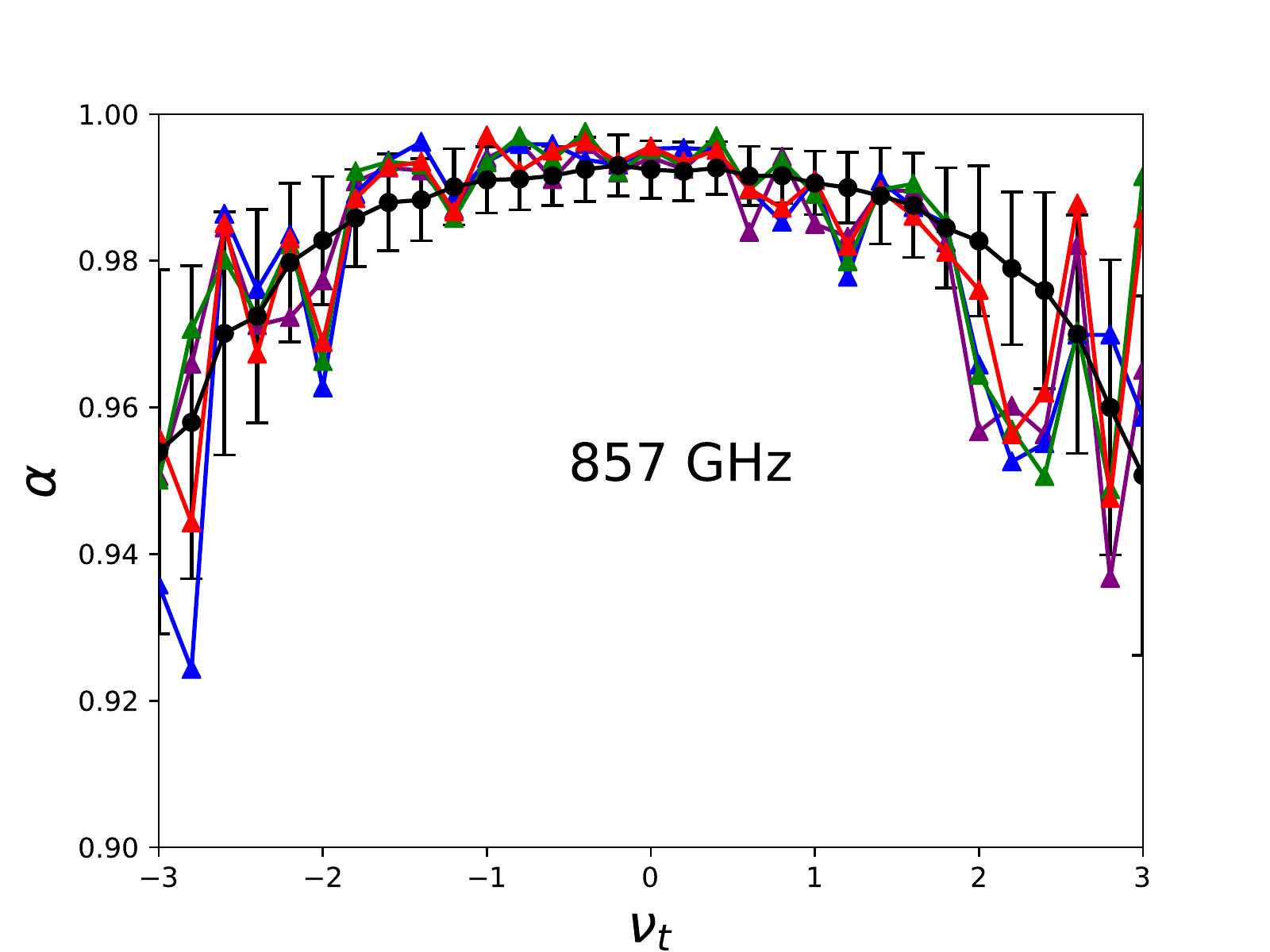}}
        
    \caption{ Comparison of $\alpha^{\rm{\, sim}}$ for beam-convolved CMB temperature maps from 100 \planck\ simulations and $\alpha^{\rm{\, data}}$ from the \planck\ data at various observation frequencies. There seems to be good agreement between the data and the simulations which include the instrumental beam effect for all frequency channels, except 30 GHz.}
    \label{fig:alpha_LFI}
\end{figure}

\subsection{Quantifying the difference between simulations and data} In order to measure the deviation of the data from the simulations, we define the quantity, $D_{\alpha}$, as follows,

\begin{equation}
 D_\alpha \equiv \left| \frac{\alpha^{\rm{\, data}}-\alpha^{\rm{\, sim}}}{\sigma_{\alpha}^{\rm{\, sim}}} \right|
\end{equation}
where, the superscripts, ``data'' and ``sim'' denote the observational data and the \planck\ \lcdm\ simulations respectively. $\sigma_{\alpha}^{\rm{\, sim}}$ is the $1\sigma$ error bar which is the standard deviation in $\alpha$, for each threshold, over the 100 simulations. $D_{\alpha}$ quantifies the deviation of the data from the simulations. A smaller value of $D_{\alpha}$ indicates good agreement between the theory and the observations. As explained in section \ref{sec:computemt}, for thresholds away from the mean, $\overline{\mathcal{W}}_1$ is averaged over fewer structures and hence $\alpha$ values calculated for these thresholds are less signficant. To quantify the level of disagreement between the data and the simulations, the $D_{\alpha}$ values should be averaged over a range of thresholds for which the $\alpha$ values are highly significant. In this work, we average the $D_{\alpha}$ values over the threshold range $-0.8$ to $+0.8$, for each dataset. The $D_{\alpha}$ values for maps smoothed to an \fwhm\ of 20' are given in Table~\ref{tab:dalpha}. If instead, we average $D_{\alpha}$ over the threshold range $-1$ to $+1$, then the disagreement between the data and the simulations is reduced, but the disagreement for the 30 GHz channel still remains significant at $\approx$1.9-$\sigma$.

\begin{table}
\resizebox{\textwidth}{!}{
\begin{tabular}{ |c|c|c|c|c|c|c|c|c|c|}
    \hline
     & & & & & & & & & \\
    $\mathcal{D}_{\alpha}$ & 30 GHz & 44 GHz & 70 GHz & 100 GHz & 143 GHz & 217 GHz & 353 GHz & 545 GHz & 857 GHz \\
     & & & & & & & & & \\
    \hline
     & & & & & & & & & \\
    COM  & 2.03  & 0.59 & 0.63  & 0.73 & 0.75 & 0.66 & 0.67 & 0.70 & 0.73 \\
     & & & & & & & & & \\
    \hline
     & & & & & & & & & \\
    \nilc\  &  1.89 & 0.82 & 0.71 & 0.63 & 0.68 & 0.75 & 0.76 & 0.75 & 0.77 \\
     & & & & & & & & & \\
    \hline
     & & & & & & & & & \\
    \sevem\  & 2.01 & 0.60 & 0.69 & 0.66 & 0.72 & 0.69 & 0.67 & 0.65 & 0.71 \\
     & & & & & & & & & \\
    \hline
     & & & & & & & & & \\
    \smica\  & 2.01 & 0.69 & 0.78 & 0.70 & 0.61 & 0.60 & 0.62 & 0.61 & 0.61 \\
     & & & & & & & & & \\
    \hline
\end{tabular}
}
\caption{$\mathcal{D}_{\alpha}$ values averaged over thresholds, $\nu_t$, from $-0.8$ to $+0.8$, for CMB temperature for various \planck\ datasets smoothed with a Gaussian beam of \fwhm=20\arcm.}
\label{tab:dalpha}
\end{table}


\section{Conclusions}

We have applied the semi-analytic expression for the Contour Minkowski Tensor  to the CMB temperature maps from the \planck\ 2015 data release \cite{PLA}. The ratio of the eigenvalues of the CMT, denoted by $\alpha$, gives a measure of the SI of the field. We have computed $\alpha$ for the \smica\ CMB temperature map and compared it with those computed from the \ffp\ \smica\ CMB simulations, which include the effect of residual foregrounds and noise. We find no significant difference between the alignments of structures found in the CMB data, and those in the CMB simulations.

Further, we have compared $\alpha$ for the individual frequency \planck\ CMB maps, for each of the four cleaning methods adopted by Planck, with the corresponding $\alpha^{\rm sim}$ computed from the simulations given in the \planck\ 2015 data release. The simulations include the instrumental beam effect. For the individual frequency maps, we find good agreement between the simulations and the observations, $< 1\sigma$, across all the frequency channels of \planck, except the 30 GHz channel. In the case of 30 GHz, we find $\approx 2\sigma$ difference in the $\alpha$ values calculated from the observed data and those calculated from the simulations, when the maps are smoothed to an \fwhm\ of 20\arcm. This mild disagreement most likely originates from the inaccurate estimation of the instrumental beam at 30 GHz in the $\ensuremath{\tt FFP9}$ simulations. It is important to have a better understanding of the source of this disagreement before making further inferences from the CMB data.

The results obtained in our paper agree well with the conclusions of \cite{Vidhya:2016} for temperature data. However, the level of agreement between observed data and simulations is higher. This confirms that the stereographic projection had a signficant contribution to the disagreement between data and simulations, in the results of \cite{Vidhya:2016}. The authors also found a 4$\sigma$ deviation from SI in the \planck\ $E$ mode polarization data released in 2015, in \cite{Vidhya:2016}. It is possible that this difference is also caused by the use of stereographic projection. We plan to extend our work to the \planck\ $E$ mode polarization data when the full dataset becomes available. It will clarify whether the 4$\sigma$ disagreement was a consequence of the stereographic projection used in the analysis, or a real disagreement between observations and \lcdm\ simulations. It is therefore, an important test of the \lcdm\ model.

\acknowledgments
We acknowledge the use of the \Hydra\ cluster at the Indian Institute of Astrophysics. Some of the results in this paper have been obtained by using the \camb\ \cite{Lewis:2000ah, cambsite} and \healpix\ \cite{HPX,Gorski:2005} packages. We acknowledge the use of observational data obtained with \planck, and ESA science missions with instruments and contributions directly funded by ESA Member States, NASA and Canada.


\end{document}